\documentstyle[12pt,aasms4]{article}
\newcommand{\etal}{{\it et al.} }
\newcommand{\asca}{{\it ASCA} }

\newcommand{\exosat}{{\it EXOSAT} }
\newcommand{\ginga}{{\it Ginga} }

\newcommand{\mcg}{MCG $-$6$-$30$-$15 }
\newcommand{\pks}{PKS~0637$-$752 }

\begin{document}

\title{A PECULIAR EMISSION-LINE FEATURE IN THE X-RAY SPECTRUM
OF THE QUASAR \pks}

\author{T. Yaqoob\altaffilmark{1,2},
I. M. George\altaffilmark{1,2},
T. J. Turner\altaffilmark{1,2},
K. Nandra\altaffilmark{1,3},
A. Ptak\altaffilmark{4},
P. J. Serlemitsos\altaffilmark{1}}

\vspace{3cm}

\altaffiltext{1}{Laboratory for High Energy Astrophysics, 
NASA/Goddard Space Flight Center, Greenbelt, MD 20771, USA.}
\altaffiltext{2}{Universities Space Research Association}
\altaffiltext{3}{NAS/NRC Research Associate}
\altaffiltext{4}{Carnegie Mellon University, Department of Physics, Pittsburgh, PA 15213 }

\vspace{5cm}

\begin{center}
{\it Accepted July 1998 for publication in Astrophysical Journal Letters}
\end{center}

\begin{abstract}
We report the results from an \asca observation of the 
high-luminosity, radio-loud quasar \pks (redshift 0.654), covering the 
0.8--15~keV band in the quasar-frame.
We find the source to have a 
luminosity $\sim10^{46}\ {\rm erg\ s^{-1}}$ in the
2--10~keV band, 
a factor of $\sim 3$ lower than during a previous \ginga observation.
The continuum appears to be well modeled by a simple power-law with 
$\Gamma = 1.64 \pm 0.07$, with no evidence for absorption 
by material intrinsic to the quasar, or Fe-K emission
(with an equivalent width $\lesssim 80$~eV at 90\% confidence).
However we do find evidence for a narrow emission line at an
energy $1.60 \pm 0.07$ keV and equivalent width 
$59^{+38}_{-34}$ eV (both in the quasar frame).
Line emission at these energies has not been observed in any
other active galaxy or quasar to date. 
We reject the possibility that this line is 
the result of instrumental artifacts, and briefly explore 
possible identifications.
\end{abstract}

\section{INTRODUCTION}

The picture emerging from many X-ray studies over
the last couple of decades of 
active galactic nuclei (AGNs) is that emission 
features in the X-ray spectra 
present in low-luminosity objects (Seyfert 1 galaxies) 
become scarce in AGNs 
with 2--10 keV intrinsic luminosity exceeding $\sim 10^{45}$
$ \rm erg \ s^{-1}$ (synonymously, quasars). These trends
are discussed at length in Nandra \etal (1997a, 1998) and Reeves \etal (1997;
and references therein). The transition to a featureless 
X-ray power-law continuum 
(except for possible line-of-sight absorption at low energies) 
in the high luminosity AGNs, especially radio
loud sources, is not fully understood but may be related to the complete
ionization of matter responsible for emission-line features and/or
beaming of the X-ray continuum swamping out emission-line features.   

The radio-loud quasar
\pks 
($z=0.654$, Hunstead, Murdoch \& Shobbrook 1978)
has been observed by the {\it Einstein} IPC (Wilkes \& Elvis 1987;
Zamorani \etal 1981, 1984; Elvis and Fabbiano 1984), \exosat
(Lawson \etal 1992; Comastri \etal 1992; Saxton \etal 1993), 
and \ginga (Williams \etal 1992;
Lawson and Turner 1997).
These X-ray observations had poor energy resolution and yielded
only continuum parameters for a simple power-law plus neutral absorber
model. The most tightly constrained measurements of the
slope were from \ginga
with a photon index of $\Gamma \sim 1.6-1.9$. The column density,
$N_{H}$, has not been well-constrained, although Lawson and Turner
(1997) claim a change from
$12 \times 10^{21} \ \rm cm^{-2}$ to $74 \times 10^{21} \ \rm cm^{-2}$
(intrinsic) between two \ginga observations. The {\it Einstein}
observations were under-exposed and so fluxes were uncertain due
to the poor spectral constraints. Nevertheless, the non-detection
of \pks in one MPC observation and a clear detection in another MPC
observation is a clear indicator of a variable continuum (Elvis and
Fabbiano 1984). \ginga still provided the most reliable 2--10 keV flux,
$\sim 9 \times 10^{-12} \rm \ erg \ cm^{-2} \ s^{-1}$ corresponding
to a luminosity of $\sim 2.6 \times 10^{46} \rm \ erg \ s^{-1}$
($H_{0} = 50 \rm \ km \ s^{-1} \ Mpc^{-1}$ and $q_{0}=0$ throughout).
In one \ginga observation the detection of an Fe-K emission line
at $\sim 6.4 $ keV in the quasar frame has been reported with
equivalent width $\sim 100$ eV but the result was marginal
(see Williams \etal 1992; Lawson and Turner 1997).

In this {\it Letter} we report on the results of a new
X-ray  
observation of \pks with {\it ASCA}
({\it Advanced Satellite for Cosmology and Astrophysics};
see Tanaka, Inoue \& Holt, 1994). 
With some surprise we have discovered an emission-line
in the object centered at $1.60 \pm 0.07$ keV in the
quasar frame. What is even more surprising is that an emission-line
at this energy has not been observed in {\it any} AGN at all, irrespective
of luminosity.

\section{THE \asca DATA}

\asca observed \pks in 1996 November 10--11 for a duration
of $\sim 120$ ks. 
Since it is important to demonstrate that our 
main observational result in this paper, namely the discovery of an
unidentified emission line, is not an
instrumental artefact, we describe the data analysis in some
detail.   
The reader is referred to Tanaka \etal (1994)
for details of the
instrumentation aboard {\it ASCA}. The two Solid State
Imaging Spectrometers (SIS),
hereafter SIS0 and SIS1, with
a bandpass of $\sim 0.5-10$ keV, were operated in 1-CCD FAINT and BRIGHT
modes. The two Gas Imaging Scintillators (GIS),
hereafter GIS2 and GIS3, with a bandpass
of $\sim 0.7-10$ keV, were operated in
standard PH mode. 
The SIS FAINT and BRIGHT mode data were combined and
the SIS energy scale was fixed using a prescription based on
measurements of Cas A (Dotani \etal 1997). This corrects for 
the continuing decline in the CTI of the CCDs with time
and is based on interpolation or extrapolation of the Cas A 
measurements. The last available Cas A measurements were made just
three months before the \pks observation and since the CTI changes
very slowly with time, the extrapolation results in a systematic
uncertainty in the gain
which is much less than $1\%$ at 1 keV (Dotani \etal 1997). 
Version 1.1 of the SIS response matrix generator was used in the
analysis. However,
no corrections were made for any possible offsets,
fluctuations and distortion of the CCD dark
level distribution (the latter also known as the `RDD' effect).
Such corrections (which themselves are subject to uncertainty) 
can only be applied to FAINT mode data and
for 1-CCD mode their effects will only be noticeable
for very high statistical quality data.  
Considering the signal-to-noise of our data
these effects can safely be neglected, since there are not enough photons
to utilize the full energy resolution of the SIS and the overall uncertainty
in the energy scale is less than $\sim 50$ eV (Dotani \etal 1997).  
More importantly, as a calibration control, we will examine data for
two other AGN observed by \asca shortly before and shortly after
\pks (\S 4).
 
For the purpose of assessing the spectral results below, we note that
spectral fits to GIS data from the Crab 
have systematics $<3\%$ over the range 0.7--10 keV 
with the current calibrations (Fukuzawa, Ishida, and Ebisawa 1997). 
Below $\sim 0.7$ keV the
GIS efficiency is greatly reduced and the calibration uncertain.
In the analysis below we include the GIS data down to 0.5 keV for the
sake of continuity but this does not affect our results because the
data are dominated by statistical errors at low energies. From $\sim 0.7$
keV to $\sim 6$ keV the SIS
calibration is generally consistent with the GIS, 
and {\it BeppoSAX LECS } and {\it MECS} to $\sim 10\%$
or better (see Grandi \etal 1997;
Orr \etal 1998). Below $\sim 0.7$ keV the uncertainty is larger,
especially between $\sim 0.5-0.6$ keV. However, our result is not
sensitive to the absolute calibration at these energies and,
in any case, we will
compare the SIS data for \pks with data
from other sources, taken shortly before and after the \pks observation.

Data were screened so that accepted events satisfied the following criteria:
(i) data were taken outside the SAA; (ii) the time since or before passage
through the SAA or a satellite day/night transition is $>50$ s; (iii)
the elevation angle to Earth is $>5^{\circ}$; (iv) the magnetic 
cut-off rigidity (COR) is $>7\ {\rm GeV/c}$; the deviation of the satellite from
the nominal pointing position was $<0.01^{\circ}$; (v) the SIS parameters
measuring active CCD pixels registered $<100$ active pixels per second,
and (vi) the radiation-belt monitor registered $<500$ ct/s.
Hot and flickering pixels in the SIS
were removed.
Screening resulted in net `good' exposure times in the range $\sim 
45.5$--49.1 Ks for the four instruments.

Images were accumulated from the
screened data. The \asca images, as well as a {\it PSPC}\ 
image obtained from archival data (a 1.3 ks observation made
in 1992 May 26),  were analysed for possible nearby contaminating
sources. No such sources were detected.

Source events were extracted from circular regions with radii
of 4' for the SIS and 6' for the GIS. Background events were 
extracted from off-source regions. As a check, background spectra
were also made from a 1-CCD blank-sky observation which resulted
from the non-detection of the quasar IRAS 15307+3252 (observed 1994
July). These background data were subject to identical selection criteria
and used to check the invariance of the spectral results for
\pks described below.
  
The \asca lightcurves do not show significant variability,
the highest excess variance (e.g. see Nandra \etal 1997b) 
we obtain is from SIS0 and has the value
$(6.9 \pm 5.8) \times 10^{-3}$ for 128 s bins.
This is to be compared
with  $0.111 \pm 0.005$ we obtain 
for the highly variable Seyfert 1 \mcg from a 4 day \asca
observation.
The count rates in the accumulated background-subtracted
spectra ranged from $\sim 0.22$--0.41 ct/s. Spectra were binned
to have a minimum of 20 counts per bin in order to utilize 
$\chi^{2}$ as the fit statistic. 

\section{SPECTRAL FITTING RESULTS}

We fitted spectra from the four \asca instruments simultaneously
in the range 0.5--9.5 keV with a simple power-law plus 
cold, neutral absorber model. 
A total of two interesting parameters were involved (the
photon index,  $\Gamma$, and column density, $N_{H}$), plus
four independent instrument normalizations. (The deviation
of any of the four normalizations from their mean was less than 7\%).
The results are shown in Table 1.
The best-fitting photon index,  $\Gamma = 1.64 \pm 0.07$, is consistent
with typical historical values (\S 1).
The column density, $N_{H}$, is consistent with the Galactic
value of $9.1 \times 10^{20} \ \rm cm^{-2}$ obtained from
Dickey and Lockman (1990). 
The instrument-averaged 
{\it observed} 0.5-2 keV and 2--10 keV fluxes are
1.3 and $3.3 \times 10^{-12} \rm \ erg \ cm^{-2} \ s^{-1}$ 
respectively. The 0.5--10 keV luminosity in the source frame
is $1.1 \times 10^{46} \rm \ erg \ s^{-1}$. 
Thus the source has
dimmed by almost a factor of 3 relative to the \ginga observations
made in 1998 July 4 and 1990 August 23 (Lawson and Turner 1997).
However, the flux is consistent (within systematic uncertainty) with
an \exosat observation made in 1984 September 8 (Comastri \etal 1992).

Although the above simple continuum
model provides an adequate fit, inspection of the ratios of data to
model (Figure~1) reveal a statistically significant `hump' at $\sim 1$ keV
(observed frame) in both SIS0 and SIS1. The GIS data are ambiguous
since the effective area is falling rapidly and the energy resolution
is worse than the SIS.

We repeated the four-instrument fits
with the addition of a simple Gaussian to model the apparent emission-line
feature. Three extra parameters were
involved, namely the line center energy, $E_{c}$; the
Gaussian intrinsic width, $\sigma$, and the line
intensity, $I$. {\it Note that we will refer to all three
parameters, plus the equivalent width (EW), in the quasar frame}.
Initially we allowed all three parameters to float and the results
are shown in Table 1. The effect on the
continuum parameters is negligible
($< 1\%$ in both $\Gamma$ and $N_{H}$).
It can be seen that the line is not resolved.
The results from repeating the fit with $\sigma$ fixed at 0.01 keV
(narrow line) 
are also given; 
the difference in $\chi^{2}$ is completely negligible (0.1) so we use
this fit to derive bounds on $E_{c}$ and EW. 
Note that the decrease in $\chi^{2}$ compared to the
model without an emission line is $>14$ for the addition of
two free parameters so the line feature is detected at
a confidence level much greater than 99\%. It is appropriate here to use 
a $\Delta \chi^{2} = 4.61$ criterion to derive 90\% confidence errors
on the best-fitting line parameters since we want to know the bounds
on $E_{c}$ and EW, independently of $\Gamma$ and $N_{H}$ (e.g. see Yaqoob 1998).
We obtain $E_{c} = 1.60 \pm 0.07$ keV and $\rm EW = 59^{+38}_{-34}$ eV. 

We also searched for Fe-K line emission 
by adding a Gaussian 
with center energy fixed at various values in the range 6.4--6.97 keV 
(with the low-energy line removed). Neither a narrow ($\sigma =0$) or
broad ($\sigma=0.5$) Gaussian component 
was statistically required, the 90\% one-parameter
upper limits on the EW being 41 and 82 eV respectively.

All the above spectral results were repeated with the alternative
background spectra made from a blank field, as described in \S 2, and 
the results were confirmed to be robust to the background used. 

\section{THE REALITY OF THE EMISSION-LINE FEATURE}

Since an emission-line feature like that reported
above has never before been observed in a quasar
(or indeed any active galaxy), we sought explanations which did not
require the line to be intrinsic to the quasar. 

To check whether the SIS calibration is responsible we examined
SIS data for 3C 371 (a quasar with $z=0.051$, observed five days before
\pks) and for Pictor A (a radio galaxy with $z=0.034$, observed fifteen
days after $PKS 0637-752$ see also Eracleous and Halpern, 1998). 
Figure 2 shows the ratios of data to  a
simple power-law plus absorber model
(both SIS0 and SIS1) for all three sources
fitted over the 0.5--9.5 keV range, but zoomed
in the range 0.5--1.5 keV
(observed frame). It can be seen that the emission-line feature
in \pks is not present in 3C 371 or Pictor A.
 
We have already stated in \S 2 that there are no contaminating sources
in the \asca or {\it PSPC} images. The emission line cannot be a
background feature (for example from Galactic diffuse emission)
because the background from nearby regions in the same
field and at the observed energy of the emission line in the SIS
lies at least a  factor 12 below the on-source spectrum.
It is unlikely that the emission line is produced in the intervening
galaxy at $z=0.469$ (Elvis and Fabbiano 1984) since the
galaxy would have to be extremely luminous (almost as luminous as the quasar is
thought to be). We also examined the {\it PSPC} spectra. These data show 
residuals to a power-law model which suggest the presence of a line 
like that observed by {\it ASCA}. However, the PSPC data do not allow 
the presence of the line to be statistically confirmed, as the 
energy resolution is extremely poor and 
signal-to-noise is very low because the exposure time is only
1.3 ks. 

We conclude that the emission line at an energy of 1.60 keV in the
rest-frame of \pks is not an instrumental artifact and is almost certainly 
associated with the quasar itself.

\section{DISCUSSION AND CONCLUSIONS}

We have reported the results of a new X-ray observation of \pks
with {\it ASCA}. 
Although the luminosity had diminished by
a factor of $\sim 3$ compared to previous \ginga observations
(six and eight years earlier), the
\asca data are the most sensitive to date in the 0.5--10 keV band ($\sim
0.8-15$ keV in the quasar frame). However, the \asca
luminosity is consistent with an \exosat observation
made twelve years earlier. It is possible that the larger
field of view of \ginga was confused with another source
which may have contributed to the higher luminosity so
the apparent variability should be treated with caution.
We find the continuum to be well
modeled by a simple power-law with $\Gamma = 1.64 \pm 0.07$ and Galactic
absorption. 
We find no evidence for Fe-K emission although the
data allow a line with moderate equivalent width
(90\% confidence quasar-frame upper limits 
in the range 41--82 eV, 
corresponding to line intrinsic widths in the range 0--0.5 keV
respectively).

Our most important result 
from the \asca observation is the discovery of a narrow emission line at an 
energy $1.60 \pm 0.07$ keV. We have presented strong evidence that this is 
not an instrumental artefact and is likely to be associated with \pks itself. 
With conservative assumptions for systematic errors, we find the line center 
to lie in the range 1.48--1.72 keV and to have an EW of $59^{+38}_{-34}$ eV in 
the quasar frame. 
Line emission in this energy band has not been observed in any 
other AGN to date, and its identification is not at all obvious. 
The allowed range of energies encompasses transitions from 
Mg{\sc xi}--{\sc xii}, Fe{\sc xxii}--{\sc xxvi} and 
Ni{\sc xix}--{\sc xxviii} (assuming no bulk motion).
Highly ionized gas is commonly seen in absorption in 
low luminosity Seyfert 1 galaxies (e.g. Reynolds 1997; George \etal 1998)
and in emission in some Seyfert 1 (below) and many 
Seyfert 2 galaxies (e.g. Turner \etal 1997).
However theoretical models of both optically-thick and optically-thin 
gas predict the stronger lines elsewhere in the 
spectrum, particularly due to a blend of O, Ne and Fe lines 
$\lesssim 1.5$~keV
(e.g. see Zycki \etal 1994; Kallman \etal 1996; Netzer 1996, and 
references therein).
Indeed the presence of emission lines at such energies 
was suggested in 
{\it Einstein} SSS observations (e.g. Turner \etal 1991)
and have been detected in 
the \asca\ spectra of
NGC 4151 (Figure 1 in Yaqoob \etal 1995),
PG~1244+026 (blend at $\sim 1$~keV, Fiore \etal 1998), 
and Ton~S~180 (blend at $\lesssim 1$~keV, Turner, George, Nandra 1998).
The detection of O{\sc vii}(0.57~keV) emission has been claimed in the case 
the \asca\ data from NGC~3783 (George, Turner, Netzer 1995), 
Mkn 290 (Turner \etal 1996), and IC~4329A (Cappi \etal 1996),
although uncertainties in the 
calibration $<0.6$~keV make their reality less compelling.
However, none of these lines are consistent with the observational
result for \pks without an 
assumption of bulk in/outflow. 
The emission feature closest in energy to that reported in \pks
and which has also been observed in \asca data from 
both Seyfert 2 galaxies (Turner \etal 1997) and starburst galaxies
(e.g. Ptak \etal 1997)
is the Si{\sc xiii}(1.84--1.86~keV) blend. 
However, the identification of the line observed in 
\pks with such emission not only requires an anomalous Si abundance
relative to  O, Ne, S and Fe, but redshifting of the new emission-line
energy by  
$\Delta z \sim0.05$--0.30
(i.e. the
line-emitting matter must be {\it in}falling at 
velocities $\gtrsim 0.05c$) with respect to the 
optical emission line gas which has 
$z=0.654$ (Hunstead \etal 1978).
Future observations of this
intriguing X-ray emission-line feature
at high spectral resolution 
with {\it AXAF}, {\it XMM} and {\it Astro-E}, will clearly  
help to shed light on its identity and origin.

The authors thank the \asca mission operations team at ISAS, Japan,
and all the instrument teams 
for their dedication and hard work in making these \asca observations 
possible. This research made use of the HEASARC archives at the
Laboratory for High Energy Astrophysics, NASA/GSFC.

\newpage

\begin{deluxetable}{lccc}
\tablecaption{Spectral Fits to \pks}
\tablecolumns{4}
\tablewidth{0pt}
\tablehead{
\colhead{Parameter} & \colhead{No line} & \colhead{broad line} &
                \colhead{narrow line}
}

\startdata

$\Gamma$
        & $1.64^{+0.07}_{-0.07}$        & $1.63^{+0.08}_{-0.07}$
        & $1.63^{+0.08}_{-0.07}$ \nl
$N_{H}$ ($10^{20} \ \rm cm^{-2}$)
        & $8.9^{+3.2}_{-3.1}$           & $10.2^{+3.8}_{-3.2}$
        & $10.1^{+3.6}_{-3.1}$  \nl
$\sigma$ (keV)
        &                               &  $0.07^{+0.24}_{-0.07}$
        & 0.01 (FIXED) \nl
$E_{c}$ (keV)
        &                               & $1.60^{+0.12}_{-0.14}$
        & $1.60^{+0.07}_{-0.07}$ \nl
$I$
        &                               & $4.9^{+7.0}_{-3.5}$
        & $4.3^{+2.9}_{-2.5}$ \nl
EW (eV)
        &                               & $66^{+94}_{- 47}$
        & $58^{+39}_{-34}$ \nl
$\chi^{2}$
        & 568.5                         & 553.8
        & 553.9  \nl
degrees of freedom
        & 530                           & 527
        & 528 \nl
\tablecomments{The continuum model used is a simple power law
plus absorber with
the photon index, $\Gamma$, and column density, $N_{H}$, floating.
The absorber is placed at $z=0$. The emission line parameters
are all referred to in the quasar frame ($z=0.654$); these
are the intrinsic width, $\sigma$, the center energy, $E_{c}$,
the intensity ($\rm 10^{-5} \rm \  photons \  cm^{-2} \ s^{-1} $),
and the equivalent width (EW).
Errors are 90\% confidence for two interesting parameters
($\Delta \chi^{2} = 4.61$)
for the continuum and for the line when it is narrow.
Otherwise the errors are
90\% confidence for three interesting parameters ($\Delta \chi^{2} =
6.251$) for the line parameters when the intrinsic line width
is floating.}
\enddata
\end{deluxetable}

\newpage
\section*{Figure Captions}

\par\noindent
{\bf Figure 1} \\
Ratio of the \asca data for \pks ($z=0.654$)
to the best-fitting power-law plus
absorber model. Notice the significant residuals in the
SIS0 and SIS1 data, corresponding to
$\sim 1.6$ keV in the quasar frame (see text).

\par\noindent
{\bf Figure 2} \\
Ratio of SIS data to best-fitting 
power-law plus absorber models for 3C 371 ($z=0.051$),
\pks ($z=0.654$) and
Pictor A ($z=0.034$). The 3C 371 were taken only five days before the
observation of \pks while the Pictor A data were taken fifteen days
after \pks. Thus calibration uncertainties in the
SIS cannot account for the emission-line feature seen in 
\pks. Note that the best-fitting parameters for 3C 371 are
$\Gamma = 1.76^{+0.06}_{-0.07}$ and $N_{H}=5.1^{+2.5}_{-2.5}
\times \rm \ 10^{20} \ cm^{-2}$ and for Pictor A they are 
$\Gamma = 1.84^{+0.02}_{-0.03}$ and $N_{H}=11.1 \pm 1.0
\rm \times \ 10^{20} \ cm^{-2}$
(errors correspond to $\Delta \chi^{2} = 4.61$).

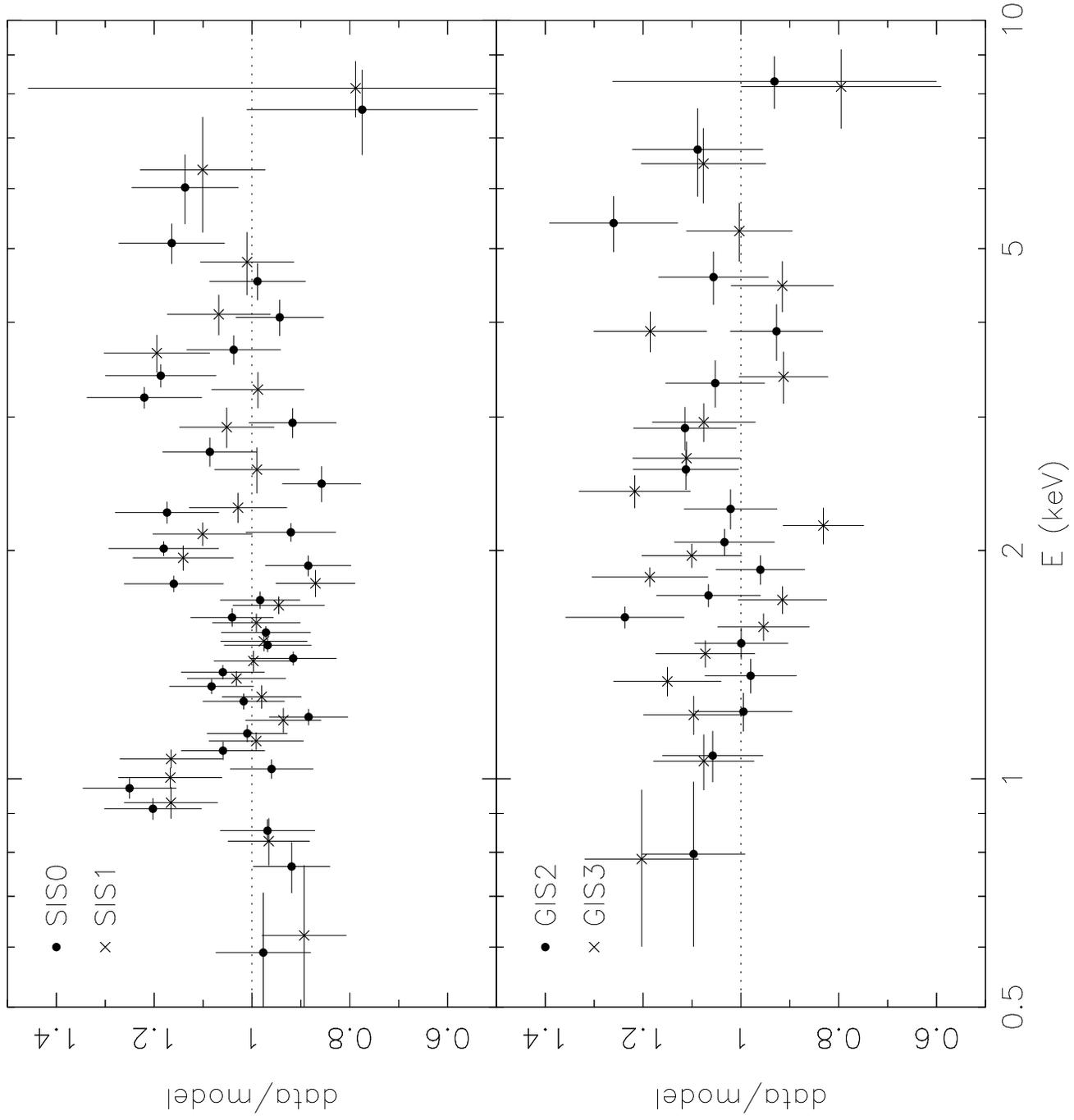
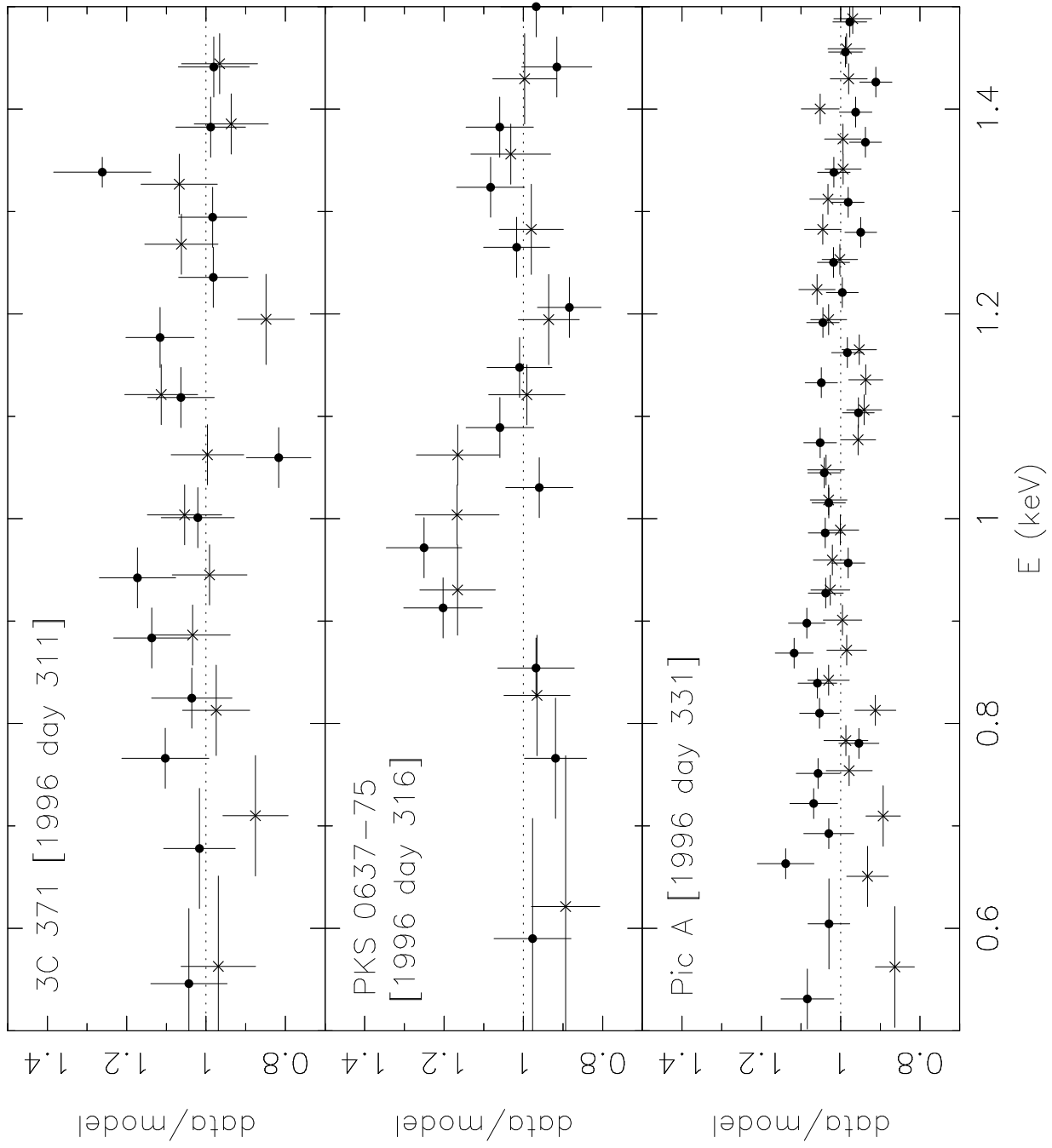
\end{document}